\documentclass[conference,english]{IEEEtran}
\IEEEoverridecommandlockouts
\usepackage{cite}
\usepackage{amsmath,amssymb,amsfonts}
\usepackage{algorithmic}
\usepackage{graphicx}
\usepackage{textcomp}
\usepackage{xcolor}
\usepackage{float}
\usepackage{graphicx}
\usepackage{setspace}
\usepackage{amssymb}
\usepackage{babel}
\usepackage{acronym}
\usepackage{units}
\usepackage{url}
\usepackage{sidecap}
\usepackage{todonotes}
\usepackage{multirow}

\acrodef{VWBump}[VWBump]{Variable Width Bump}
\acrodef{DPI}[DPI]{Diff-Pair Integrator}

\newcommand{\refeq}[1]{{Eq.~(\ref{#1})}}

\newcommand{\reffig}[1]{{Fig.~\ref{#1}}}
\newcommand{\reftab}[1]{{Tab.~\ref{#1}}}

\newcommand{\refsec}[1]{{Sec.~\ref{#1}}}
\normalsize

\def\BibTeX{{\rm B\kern-.05em{\sc i\kern-.025em b}\kern-.08em
    T\kern-.1667em\lower.7ex\hbox{E}\kern-.125em}}
\begin{document}
\acrodef{AC}[AC]{Arrenhius \& Current}
\acrodef{ANN}[ANN]{Artificial Neural Network}
\acrodef{AER}[AER]{Address Event Representation}
\acrodef{AEX}[AEX]{AER EXtension board}
\acrodef{AMDA}[AMDA]{``AER Motherboard with D/A converters''}
\acrodef{API}[API]{Application Programming Interface}
\acrodef{BP}[BP]{Back-Propagation}
\acrodef{BPTT}[BPTT]{Back-Propagation-Through-Time}
\acrodef{BM}[BM]{Boltzmann Machine}
\acrodef{CAVIAR}[CAVIAR]{Convolution AER Vision Architecture for Real-Time}
\acrodef{CCN}[CCN]{Cooperative and Competitive Network}
\acrodef{CD}[CD]{Contrastive Divergence}
\acrodef{CMOS}[CMOS]{Complementary Metal--Oxide--Semiconductor}
\acrodef{COTS}[COTS]{Commercial Off-The-Shelf}
\acrodef{CPU}[CPU]{Central Processing Unit}
\acrodef{CV}[CV]{Coefficient of Variation}
\acrodef{CV}[CV]{Coefficient of Variation}
\acrodef{DAC}[DAC]{Digital--to--Analog}
\acrodef{DBN}[DBN]{Deep Belief Network}
\acrodef{DCLL}[DECOLLE]{Deep Continuous Local Learning}
\acrodef{DFA}[DFA]{Deterministic Finite Automaton}
\acrodef{DFA}[DFA]{Deterministic Finite Automaton}
\acrodef{divmod3}[DIVMOD3]{divisibility of a number by 3}
\acrodef{DPE}[DPE]{Dynamic Parameter Estimation}
\acrodef{DPI}[DPI]{Differential-Pair Integrator}
\acrodef{DSP}[DSP]{Digital Signal Processor}
\acrodef{DVS}[DVS]{Dynamic Vision Sensor}
\acrodef{EDVAC}[EDVAC]{Electronic Discrete Variable Automatic Computer}
\acrodef{EIF}[EI\&F]{Exponential Integrate \& Fire}
\acrodef{EIN}[EIN]{Excitatory--Inhibitory Network}
\acrodef{EPSC}[EPSC]{Excitatory Post-Synaptic Current}
\acrodef{EPSP}[EPSP]{Excitatory Post--Synaptic Potential}
\acrodef{eRBP}[eRBP]{Event-Driven Random Back-Propagation}
\acrodef{FPGA}[FPGA]{Field Programmable Gate Array}
\acrodef{FSM}[FSM]{Finite State Machine}
\acrodef{GPU}[GPU]{Graphical Processing Unit}
\acrodef{HAL}[HAL]{Hardware Abstraction Layer}
\acrodef{HH}[H\&H]{Hodgkin \& Huxley}
\acrodef{HMM}[HMM]{Hidden Markov Model}
\acrodef{HW}[HW]{Hardware}
\acrodef{hWTA}[hWTA]{Hard Winner--Take--All}
\acrodef{IF2DWTA}[IF2DWTA]{Integrate \& Fire 2--Dimensional WTA}
\acrodef{IF}[I\&F]{Integrate \& Fire}
\acrodef{IFSLWTA}[IFSLWTA]{Integrate \& Fire Stop Learning WTA}
\acrodef{INCF}[INCF]{International Neuroinformatics Coordinating Facility}
\acrodef{INI}[INI]{Institute of Neuroinformatics}
\acrodef{IO}[IO]{Input-Output}
\acrodef{IPSC}[IPSC]{Inhibitory Post-Synaptic Current}
\acrodef{ISI}[ISI]{Inter--Spike Interval}
\acrodef{JFLAP}[JFLAP]{Java - Formal Languages and Automata Package}
\acrodef{LIF}[LI\&F]{Linear Integrate \& Fire}
\acrodef{LSM}[LSM]{Liquid State Machine}
\acrodef{LTD}[LTD]{Long-Term Depression}
\acrodef{LTI}[LTI]{Linear Time-Invariant}
\acrodef{LTP}[LTP]{Long-Term Potentiation}
\acrodef{LTU}[LTU]{Linear Threshold Unit}
\acrodef{MCMC}{Markov Chain Monte Carlo}
\acrodef{NHML}[NHML]{Neuromorphic Hardware Mark-up Language}
\acrodef{NMDA}[NMDA]{NMDA}
\acrodef{NME}[NE]{Neuromorphic Engineering}
\acrodef{PCB}[PCB]{Printed Circuit Board}
\acrodef{PRC}[PRC]{Phase Response Curve}
\acrodef{PSC}[PSC]{Post-Synaptic Current}
\acrodef{PSP}[PSP]{Post--Synaptic Potential}
\acrodef{RI}[KL]{Kullback-Leibler}
\acrodef{RRAM}[RRAM]{Resistive Random-Access Memory}
\acrodef{RBM}[RBM]{Restricted Boltzmann Machine}
\acrodef{ROC}[ROC]{Receiver Operator Characteristic}
\acrodef{SAC}[SAC]{Selective Attention Chip}
\acrodef{SCD}[SCD]{Spike-Based Contrastive Divergence}
\acrodef{SCX}[SCX]{Silicon CorteX}
\acrodef{SRM}[SRM]{Spike Response Model}
\acrodef{SNN}[SNN]{Spiking Neural Network}
\acrodef{STDP}[STDP]{Spike Time Dependent Plasticity}
\acrodef{SW}[SW]{Software}
\acrodef{sWTA}[SWTA]{Soft Winner--Take--All}
\acrodef{VHDL}[VHDL]{VHSIC Hardware Description Language}
\acrodef{VLSI}[VLSI]{Very  Large  Scale  Integration}
\acrodef{WTA}[WTA]{Winner--Take--All}
\acrodef{XML}[XML]{eXtensible Mark-up Language}

\title{Error-triggered Three-Factor Learning Dynamics for Crossbar Arrays}

\author{ \IEEEauthorblockN{Melika Payvand*}
\IEEEauthorblockA{Institute of neuroinformatics\\ University and ETH of Zurich, Zurich, Switzerland}
\and
\IEEEauthorblockN{Mohammed Fouda*, Fadi Kurdahi, Ahmed Eltawil}
\IEEEauthorblockA{Department of Electrical Engineering and Computer Science\\ UC Irvine, Irvine, CA 92697-2625 USA}\\
\IEEEauthorblockN{Emre O. Neftci}
\IEEEauthorblockA{Department of Cognitive Sciences and Department of Computer Science\\ UC Irvine, Irvine, CA 92697-2625 USA} }

\maketitle

\begin{abstract}
Recent breakthroughs suggest that local, approximate gradient descent learning is compatible with \acp{SNN}.
Although \acp{SNN} can be scalably implemented using neuromorphic VLSI, an architecture that can learn \emph{in situ} as accurately as conventional processors is still missing. 
Here, we propose a subthreshold circuit architecture designed through insights obtained from machine learning and computational neuroscience that could achieve such accuracy. 
Using a surrogate gradient learning framework, we derive local, error-triggered learning dynamics compatible with crossbar arrays and the temporal dynamics of \acp{SNN}.
The derivation reveals that circuits used for inference and training dynamics can be shared, which simplifies the circuit and suppresses the effects of fabrication mismatch.
We present SPICE simulations on XFAB 180nm process, as well as large-scale simulations of the spiking neural networks on event-based benchmarks, including a gesture recognition task. 
Our results show that the number of updates can be reduced hundred-fold compared to the standard rule while achieving performances that are on par with the state-of-the-art.

\end{abstract}


\section{Introduction}
The implementation of learning dynamics as synaptic plasticity in neuromorphic hardware can lead to highly efficient, lifelong learning systems \cite{Neftci18_datapowe}.
While gradient Backpropagation (BP) is the workhorse for training nearly all deep neural network architectures, it is incompatible with neuromorphic hardware because it is not spatially and temporally local \cite{Baldi_etal17_learmach}. 
Recent work addresses this problem using Surrogate Gradient (SG) learning \cite{Neftci_etal19_surrgrad}.
SGs use a differentiable surrogate network to compute weight updates in a local fashion, and formulate the updates as three-factor synaptic plasticity rules.
The SG approach reveals from first principles the mathematical nature of the three factors, and a learning dynamic that is continuous in time.
While temporal continuity is a plausible property in the brain, it leawhile being able tods to a large number of weight updates (writes) which can be energetically expensive in hardware \cite{Neftci18_datapowe}. 

Here, we demonstrate a crossbar based neuromorphic architecture that efficiently implements SG learning as a three-factor plasticity rule.
The problem of continuous updates is solved by triggering weight updates asynchronously when the error exceeds a threshold. 
We propose a subthreshold analog circuit that efficiently implements the neural dynamics and error-triggered updates.
We find that the circuits for learning and inference can be shared, which further reduces the circuit complexity, and suppresses mismatch in the peripheral circuits.
Taken together, our results demonstrate that the additional circuit complexity for efficient learning with spiking neurons is small compared to a conventional artificial neural network, and could enable efficient spatiotemporal pattern learning in memristor-based crossbar arrays.

\section{Neural Network Model}
The proposed model consists of networks of plastic integrate-and-fire neurons. 
Models here are formalized in discrete-time to make the equivalence with classical artificial neural networks more explicit.
However, these dynamics can also be written in continuous-time without any conceptual changes. The neuron and synapse dynamics are:
%
\begin{equation}\label{eq:lif_equations}
  \begin{split}
    U_i^l[n] &= \sum_j W^l_{ij} P_j^l[n] - \delta R^l_i[n], \quad S_i^l[n] = \Theta( U_i^l[n]) \\
  \end{split}
\end{equation}
\[
  \begin{split}
    P_j^l[n+1] &= \alpha_j^l P^l_{j}[n] + Q^l_{j}[n], \\
    Q_j^l[n+1] &= \beta_j^l  Q^l_{j}[n] + S^{l-1}_{j}[n], \\ 
    R_i^l[n+1] &= \gamma_i^l R^l_{i}[n] + S^{l}_{i}[n]. \\
  \end{split}
\]
where $U_i^{l}[n]$ is the membrane potential of neuron $i$ at layer $l$ at time step $n$, $W^l$ is the synaptic weight matrix between layer $l-1$ and $l$, and $S_i^l$ is the binary output of this neuron. 
The step function $\Theta$ is the step function, \emph{i.e.} ($S_i^l[n]=1$) when $U_i^l[n]=0$.
The constants $\alpha_j^l$, $\gamma_j^l$ and $\beta_j^l$ capture the decay dynamics of the membrane potential $U_i^l$, the refractory (resetting) state $R_i^l$ and the synaptic state $Q_i^l$ and can be related to time constants in leaky integrate-and-fire neurons.
The dependency of the time constants on $j$ and $l$ takes into account the circuit-to-circuit variability due to fabrication mismatch.
States $P$ and $Q$ describe the traces of the membrane and the current-based synapse, respectively. 
$R$ is a refractory state that resets and inhibits the neuron after it has emitted a spike, and $\delta$ is the constant that controls its magnitude.
Note that \refeq{eq:lif_equations} is equivalent to a discrete-time version of the \ac{SRM}$_0$ with linear filters \cite{Gerstner_Kistler02_spikneur}.
This \ac{SNN} and the ensuing learning dynamics can be transformed into a standard binary neural network by setting all $\alpha=0$, replacing all $P[n]$ with $S[n-1]$ and dropping $Q$ and $R$.

\section{Surrogate Gradient Learning}
Assuming a global cost function $\mathcal{L}$, the gradients with respect to the weights in layer $l$ are formulated as three factors 
\begin{equation}\label{eq:loss}
\frac{\partial}{\partial W_{ij}^{l}} \mathcal{L} = 
\frac{\partial}{\partial S_{i}^{l}}\mathcal{L}
\frac{\partial}{\partial U^{l}_{i}}S^{l}_i  
\frac{\partial}{\partial W^{l}_{ij}}U^{l}_i
\end{equation}
The rightmost factor 
describes the change of the membrane potential changes with the weight $W_{ij}^l$.
This term is equal to $P_j^l[n] - \frac{\partial}{\partial W^{l}_{ij}}R^l_i[n]$ for the neuron defined by \refeq{eq:lif_equations}.
The term with $R$ involves a dependence of the past spiking activity of the neuron, which significantly complicates the learning dynamics. 
Fortunately, this dependence can be ignored during learning without empirical loss in performance \cite{Zenke_Ganguli17_supesupe}. 
The middle factor
is the change in spiking state as a function of the membrane potential, \emph{i.e.} the derivative of $\Theta$.
$\Theta$ is non-differentiable but can be replaced by a smooth sigmoidal or piecewise constant function in the learning rule \cite{Neftci_etal19_surrgrad}.
Our experiments make use of a piecewise linear function, such that middle factor becomes the box function: $\frac{\partial}{\partial U^{l}_{i}}S^{l}_i := B(U_i) = 1$ if $u_-<U_i<u_+$ and $0$ otherwise.
The leftmost factor describes how the change in the spiking state affects the loss.
It is commonly called the local error (or the ``delta'') and is typically computed using gradient BP.
We assume for the moment that these local errors are available and denote them $err_i^l$, and revisit this point in \refsec{sec:local_error}.
The weight updates become:
\begin{equation}\label{eq:spiking_neuron_rule}
  \Delta W_{ij}^l = - \eta \frac{\partial}{\partial W_{ij}^{l}} \mathcal{L} = - err_i^l P_j^{l} \text{, if $u_-<U_i<u_+$},
\end{equation}
where $\eta$ is the learning rate.
\subsection{Bipolar Error-triggered Learning}
For most interesting cost functions, errors must be computed extrinsically and communicated to the neuron.
To make this communication efficient, we encode errors using bipolar events as follows:
\begin{equation}\label{eq:error-coding-neurons}
    E^l_i = sign(err_i^l)[|err_i^l|-\theta]^+
\end{equation}
where $\theta \in \mathbb{R}$ is a constant or slowly varying error threshold and $[\cdot]^+$ is the recitifed linear function. 
Using this encoding, the parameter update rule becomes:
\begin{equation}\label{eq:binary_neuron_rule}
  \Delta W_{ij}^l = - \theta E^l_i P_j^{l} B(U_i^l) 
\end{equation}
where $\theta$ is called the stop-learning threshold ($\eta$ was folded into $\theta$).
Thus, an update takes place on an error of magnitude $\theta$ and if $B(U_i^l)=1$.
The sign of the weight update is $-E^l_i$ and its magnitude $\theta P_j^{l}$. 
Provided that the layer-wide update magnitude can be modulated proportionally to $\theta$, this learning rule implies two comparisons and an addition (subtraction).

\subsection{Local Losses and Local Errors}\label{sec:local_error}
Up to now, we have side-stepped the calculation of $err[n]^l_i$.
If $l$ is not the output layer, then computing this term requires solving a deep credit assignment problem. 
Gradient BP can solve this, but is not compatible with a physical implementation of the neural network \cite{Baldi_etal17_learmach}.
Several approximations have emerged recently to solve this, such as feedback alignment \cite{Lillicrap_etal16_randsyna, Neftci_etal17_evenranda, N-kland16_direfeed}, and local losses defined for each layer \cite{Mostafa_etal17_deepsupe,Kaiser_etal18_synaplas,Nkland_Eidnes19_traineur}.
For classification, local losses can be local classifiers (using output labels) \cite{Mostafa_etal17_deepsupe}, and supervised clustering, which perform on par and sometimes better than BP in classical ML benchmark tasks \cite{Nkland_Eidnes19_traineur}.
For all experiments used in this work, we use a layer-wise local classifier using a mean-squared error loss defined as $\mathcal{L}_i^l = ||\sum_{k=1}^C J^l_{ik} S^l_k-\hat{y}_k||_2$, 
where $J^l_{ik}$ is a random, fixed matrix, $\hat{y}_k$ are one-hot encoded labels, and $C$ is the number of classes.
Because the gradients of $\mathcal{L}_i^l$ involve backpropagation, we train through feedback alignment using another random matrix $H^l$ \cite{Lillicrap_etal16_randsyna} whose elements are equal to $H_{ij}^l = J_{ij}^{l,T}\omega_{ij}^l$ with Gaussian distributed $\omega_{ij}^l \sim N(1,\frac12)$.
Weight updates are achieved through stochastic gradient descent (SGD). 
We note that our approach is agnostic to the used loss function as long as there is no temporal dependency (i.e. $L[n]$ does not depend on variables in time step $n-1$).

\subsection{Hardware Realization with Memristor Crossbar Arrays }
\label{sec:moh}

The emerging technologies, such as memristors (or RRAMs), phase change memory, and spin transfer torque-RAM in addition to other MOS realizations such as floating gate transistors, assembled as crossbar array enable the VMM operation to be completed in a single step. This is unlike other hardware solutions which requires $N\times M$ steps where $N$ and $M$ are the size of the weight matrix. 

These emerging technologies implement only positive weight (excitatory connections), however, the negative weights (inhibitory connections) are needed as well. There are two ways to realize the positive and negative weights \cite{Fouda_etal18_indecomp}; 1) balanced realization where two devices are needed to implement the weight value.
The weight value is stored in the devices' conductances where $w=G^+-G^-$.
If the $G^+$ is greater/less than $G^-$, it represents positive/negative weight, respectively.
2) Unbalanced realization where one device is used to implement the weight value with a common reference conductance which is set to the mid value of the conductance range. Thus, the weight value is represented as $w=G-G_{ref}$. If the $G$ is greater/less than $G_{ref}$, it represents positive/negative weight, respectively. In this work, we use the unbalanced realization since it saves the area and power but with half of the dynamic range. Thus, the memristive SNN can be written as  
\begin{equation}\label{eq:lif_memristor_equations}
    U_i^l[n] = \sum_j \left(G^{l}_{ij}-G_{ref}\right)P_j^l[n]
\end{equation}

\begin{SCfigure}
  \centering\includegraphics[width=.2\textwidth]{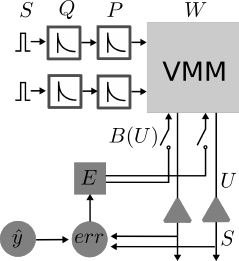}
  \caption{Architecture of the Three-Factor Error-Triggered Rule. Input spikes $S$ are integrated through $P$. The vector $P$ is then multiplied with $W$ resulting in $U$. Output spikes $S$ are then compared with local targets $\hat{y}$ and bipolar error events $E$ are fed back to each neuron. Updates are made if $u_-<U<u_+$. $R$ is omitted in this diagram }
  \label{fig:arch}
\end{SCfigure}

By following the same analysis discussed in the previous section, the conductance update model is the same as \eqref{eq:spiking_neuron_rule}. 
The general architecture of the proposed dynamics is shown in \reffig{fig:arch}.

\subsection{Inference and Learning Circuits}

\begin{figure*}[t]
  \centering\includegraphics[width=0.8\textwidth]{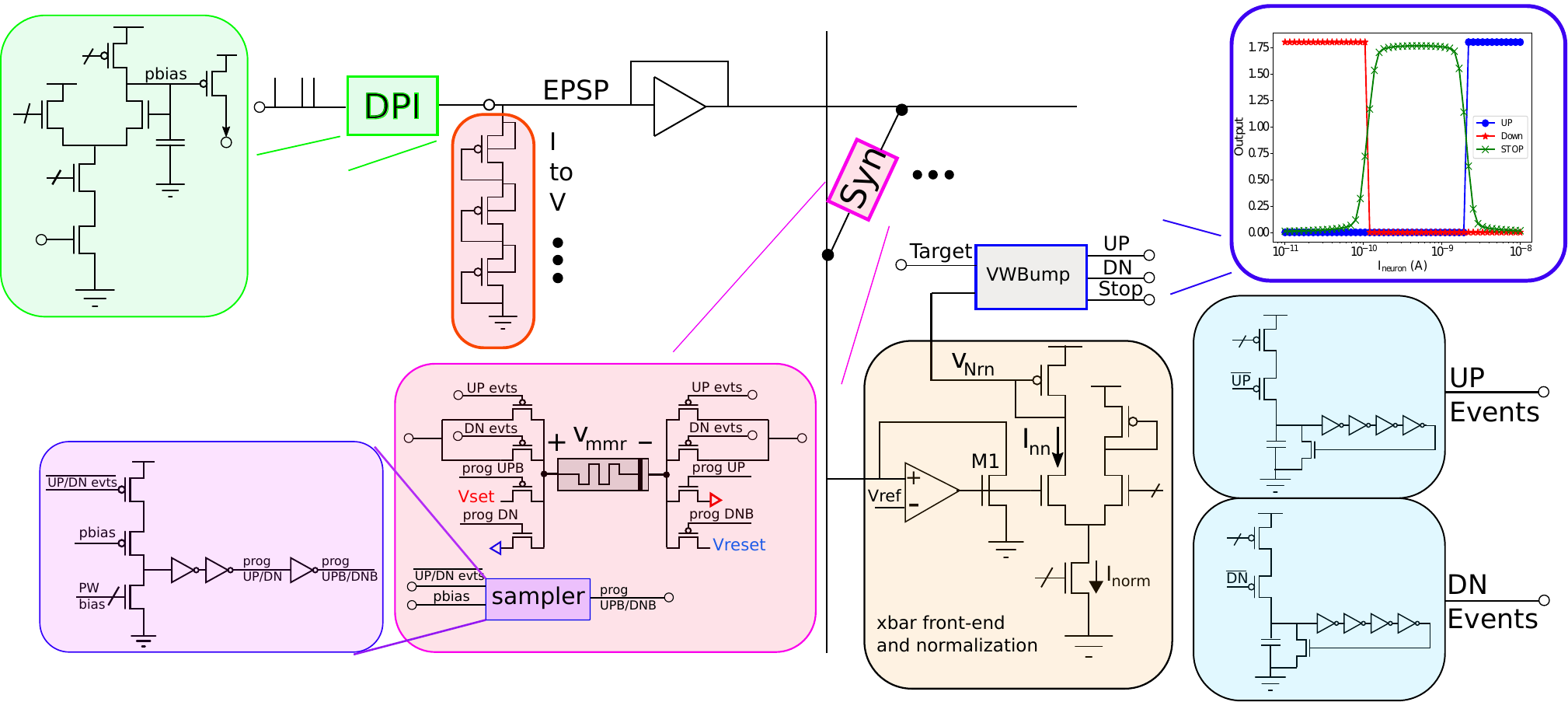}
  \caption{Details of the architecture and learning circuits. Green: DPI circuit generating P in the current form. Red: Pseudo resistors converting input current into a voltage driving the crossbar array. Pink: Synapse with the controlling switches. Purple: Sampling circuitry generating pulses to program the devices. Yellow: Crossbar front-end and normalization of the crossbar current. Dark blue: Bump circuit comparing the crossbar current to a target and generating the direction of the error. Light blue: Bidirectional neuron producing up and down events.}
  \label{fig:arch_learn}
\end{figure*}

\begin{figure}
\centering\includegraphics[width=0.45\textwidth]{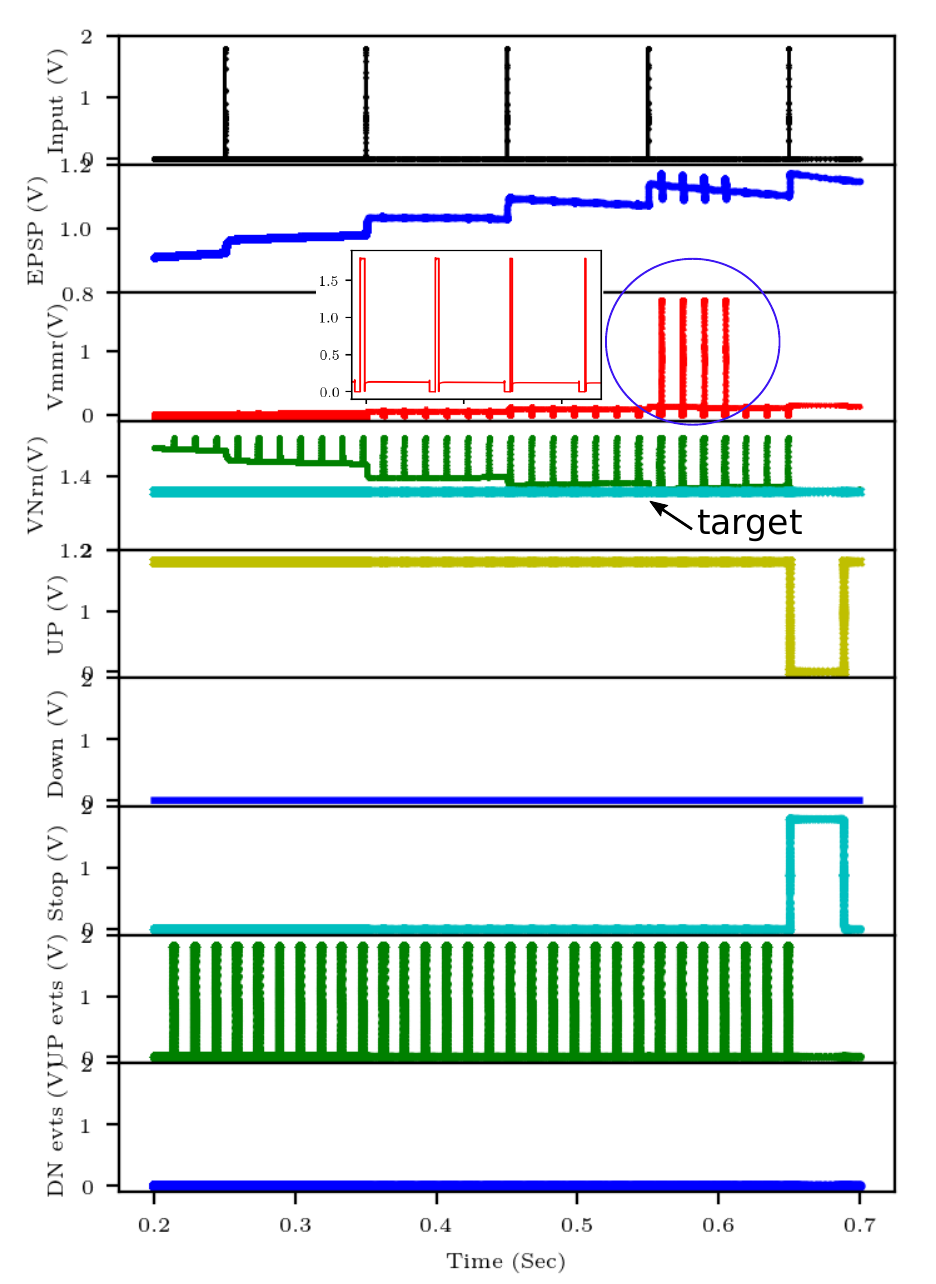}
  \caption{SPICE simulation results of the learning circuits generating the appropriate programming pulses across the memristive device ($Vmmr$) depending on the value of the $EPSP$ at the onset of the error events.}
  \label{fig:circuit_res}
\end{figure}

Our circuit implementation of the spiking neural network differs from classical ones.
Generally, the rows of crossbar arrays are driven by (spikes) and integration takes place at each column \cite{Chen_etal15_mitieffe}. 
While this is beneficial in reducing read power and mitigating sneak path problems, it renders learning more difficult because the variables necessary for learning in SNNs are not local to the crossbar. 
Instead, we use the crossbar as a vector matrix multiplication of pre-synaptic trace vectors $P^l$ and synaptic weight matrices $W^l$. 
Using this strategy, a single trace $P_i^l$ per neuron supports both inference and learning. 
Furthermore, this property means that learning is immune to the mismatch in $P^l$, and can even exploit this variation for reducing the loss.
\reffig{fig:arch_learn} depicts the details of the learning circuits in a crossbar-like architecture which is compatible with the address-event representation (AER) as the conventional scheme for communication between neuronal cores in many neuromorphic chips \cite{Deiss_etal98_pulscomm}. 
In this circuit, only $P$ is shown and $\alpha_Q=0$.
This type of architecture includes multi-T/1R \cite{Payvand_etal19_neursyst}. 
The traces $P$ are generated through a \ac{DPI} circuit which generates a tunable exponential response at each input event in the form of a sub-threshold current \cite{Bartolozzi_Indiveri07_synadyna}. 
The current is linearly converted to voltage using pseudo resistors in the I-to-V block highlighted in the red box in \reffig{fig:arch_learn}. 
The exponentially decaying voltage is buffered and drives the entire crossbar row in accordance with \eqref{eq:lif_equations}.

For every neuron, different voltages (corresponding to $P_j$) are applied to the top electrode of the corresponding memristive device whose bottom electrode is pinned by the crossbar front-end highlighted in yellow (\reffig{fig:arch_learn}).
This block pins the entire column to a reference voltage ($Vref$) and reads out the sum of the currents generated by the application of $P$s across the memristors in the column.
As a result, a voltage is developed on the gate of the M1 connected to a differential pair which re-normalizes the sum of the currents from the crossbar to $I_{norm}$. 
This ensures that the currents remain in the sub-threshold regime for the next-stage of the computation which is the ternary error generation as is specified in equation (\ref{eq:error-coding-neurons}). 
This is done through the \ac{VWBump} circuit that compares $I_{nn}$ to the target $\hat{y}$, with a stop region.
Thus, the \ac{VWBump} circuit output indicates the sign of the weight update (up or down) or stop-learning (no update).
The circuit (not shown) is based on the bump circuit \cite{Delbruck91_bumpcirc}, which consists of a differential pair for the comparison and a current correlator for the stop region, and is modified to have a tunable stop-learning region \cite{Payvand_Indiveri19_spikplas}. 
The boundaries of this region play the role of $\theta$ in (\ref{eq:error-coding-neurons}). The output of the block is plotted in the inset of \reffig{fig:arch_learn}, which shows the {Up}, {Down} and {STOP} outputs.

The {Up} and {Down} signals trigger the oscillators highlighted in blue which generate the bipolar $E_i$ events. 
According to \refeq{eq:binary_neuron_rule}, the magnitude of the weight update is $P_j$, and thus $P_j$ must be sampled at the onset of $E_i$. 
To do so, we regenerate the exponential current in the entire row by propagating {pbias} shown in the \ac{DPI} circuit block (green) and sample it by the up and down events. This is done through the sampling circuit (shown in purple) whose core consists of two PMOS transistors in series connected to the up/down events and {pbias} respectively. 
An NMOS transistor is biased to generate a current much smaller than that of the DPI and as a result, the higher the DPI current, the higher the input of the following inverter during the event pulse, and thus it takes longer for the NMOS to discharge that node.
This results in a pulse width which varies linearly with $P_j$, in agreement with \refeq{eq:binary_neuron_rule}. The linear pulse width can be approximated with multiple pulses which results in a linear conductance update (with a soft bound) in memristive devices \cite{Frascaroli_etal18_evidsoft}. \\
\reffig{fig:circuit_res} illustrates the Spectre results of the above circuits designed in XFAB 180nm process. 
With every input event, the DPI current (and therefore EPSP) undergoes a near instantaneous jump and decays exponentially.  
The EPSP is buffered and applied to the memristive device whose other side is pinned at $vdd/2$ ($0.9$V). $Vmmr$, the voltage drop across the device, follows the EPSP except for the time it is being programmed. 
$VNrn$, marked on \reffig{fig:arch_learn}, is used to mirror the normalized crossbar current ($I_{nn}$) to the bump circuit and is shown in green in \reffig{fig:circuit_res}.
In the beginning, while the EPSP is low, $I_{nn}$ is lower than the target, therefore, the {UP} output of the \ac{VWBump} circuit is high and the UP events are generated through the oscillator.
As the neuron gets closer to the target (because of the integrated input events), the STOP output of \ac{VWBump} is flipped to high and the event generation stops. 
The UP events sample the EPSP to change the synaptic weight correspondingly.
While the EPSP is low, no programming pulse is generated.
For higher values of the EPSP, the pulse width is higher and it falls as the EPSP decays.
This is highlighted in the inset of the $Vmmr$.   
Note that the memristor model (and thus the synaptic update) is not included in the circuit simulations and we are only showing the programming conditions which would cause the conductance change based on the online learning algorithm.

\section{Large-scale Simulation Experiments}
An important feature of the used learning rule is its scalability to multilayer networks with very small loss of performance compared to a standard deep neural network when using idealized dynamics.
To demonstrate this, we simulate the learning dynamics for classification in large-scale, multilayer spiking networks. 
The GPU simulations focus on event-based datasets acquired using a neuromorphic sensor, namely the N-MNIST and DVS Gestures dataset for demonstrating the learning model. 
 Both datasets were pre-processed as in \cite{Kaiser_etal18_synaplas}.
The N-MNIST network is fully connected (1000--1000--1000), while the DVS Gestures network is convolutional (64c7-128c7-128c7).
For practical reasons, the neural networks were trained in batches of 72 (DVS Gestures) and 100 (N-MNIST).
The parameters of our model are similar to that of \cite{Kaiser_etal18_synaplas} except that the time constants were randomized.
In our experiments, we used a proportional controller to adjust $\theta$ such the average error spike rate $\langle E \rangle$ remains stable.
The column writes indicates an upper bound on the number of weight writes. It is an approximate upper boundary, as the effect of $B(U)$ has not been taken into account.
These results in \reftab{tab:res} demonstrate a small loss in accuracy across the two tasks when updates are error-triggered.

\begin{table}
\begin{center}
\begin{tabular}{l l l }  
  \multicolumn{3}{l}{\bf DVSGesture}       \\
  $\langle E \rangle$ & {Error}                & Writes\\
  Cont.       & 3.82\%        & 1M \\
  50Hz       & 4.22\%        & 50k\\
  10Hz       & 6.25\%        & 10k\\ 
\end{tabular}
\begin{tabular}{l l l }  
  \multicolumn{3}{l}{\bf N-MNIST}       \\
  $\langle E \rangle$ & {Error}                & Write\\
  Cont.       & 2.3\% & 1.5M\\
  50Hz       & 2.31\% & 75k\\
  10Hz       & 2.71\% & 45k\\
\end{tabular} 
\end{center}
  \caption{Recognition Error in Idealized Spiking Neural Network Simulations Averaged over 5 Runs. }
  \label{tab:res}
\end{table}

\section{Conclusion}
In this article, we demonstrated an error-triggered learning rule that is particularly well-suited for implementation in crossbars. 
Our implementation leverages the linear property of the subthreshold dynamics, such that the memory required for computing the gradients (i.e. the synaptic traces) grows linearly with the neurons (hence one $P_j$ per input neuron). 
By updating weights asynchronously (when errors occur), the number of weight writes can be drastically reduced.
Our implementation still requires a programming circuitry (8 transistors) per synapse along with transistors which switch the memristive device in an out of the network for read/programming.
However, the transistors can take advantage of the technology scaling (in contrast to the capacitors whose area do not change as much with the scaling of the nodes).

Alternatively, one can decide to sacrifice the high density for complex synapses that include multiT-1R , to capitalize on the many other features of memristive devices (in addition to their small footprint), such as non-volatility, state-dependence, complex dynamics, and stochastic switching behavior. 


Despite of the huge benefit of the crossbar array structure, the memristor devices suffer from many challenges that might affect the performance unless taken into consideration in the training such as asymmeteric nonlinearity, precision and retention. Fortunately, online learning helps with other problems such as sneak path (i.e wire resistance) and endurance. With error-triggered learning rule, only selected devices are updated which leads to extending the lifetime of the devices and less write energy consumption. The aforementioned non-idealities will be considered in our future work.   

\section{Acknowledgements}
We acknowledge Giacomo Indiveri for fruitful discussions on the learning circuits.
This work was supported by the National Science Foundation under grant 1652159 (EON); the Korean Institute of Science and Technology 1652159 (EON); the National Science Foundation under grant 1640081 (EON, MF); and the Nanoelectronics Research Corporation (NERC), a wholly-owned subsidiary of the Semiconductor Research Corporation (SRC), through Extremely Energy Efficient Collective Electronics (EXCEL), an SRC-NRI Nanoelectronics Research Initiative under Research Task ID 2698.003 (EON, MF); and Toshiba Corporation (MP).

\bibliographystyle{IEEEtran} 
\bibliography{biblio_unique_alt}

\end{document}